\documentclass[11pt,a4paper]{article}

\usepackage{amsmath,amsthm,amssymb}
\usepackage{graphics,graphicx}
\usepackage{epsfig}
\usepackage{multicol}
\usepackage{color}
\makeatletter
\@addtoreset{equation}{section}
\makeatother

\begin{document}

\begin{center}
\textbf{\Large Thermodynamic Geometry and Type 0A Black Holes}\\
\end{center}

\begin{center}
\large{\bf Narit Pidokrajt,} ${}^{a,}$\footnote{narit@fysik.su.se} \large
\large{\bf and John Ward} ${}^{b,}$\footnote{jwa@uvic.ca}
\end{center}
\begin{center}
\emph{ ${}^a$ Fysikum, Albanova, Stockolm University \\ SE-10691 Stockholm, Sweden
\\}
\vspace{0.1cm}
\emph{${}^{b}$ Department of Physics and Astronomy, University of Victoria, Victoria,
BC \\ V8P 1A1, Canada }
\end{center}

\begin{abstract}
In this note we study thermodynamic geometry of the type 0A black hole solution in string theory using a variety of different methods 
(Ruppeiner, Weinhold and Geometrothermodynamics). Our results indicate that the curvature invariants are finite for all physical solutions, 
suggesting that there is no phase transition. It is also found that the cutoff of the entropy, which is the singular limit of the theory, 
appears geometrically in the Weinhold picture as the thermodynamic cone itself. 

\end{abstract}
\section{Introduction}
The study of black hole thermodynamics has been an intensely active subject for the
past few decades. The fact that such objects emit Hawking radiation has been at the
heart of recent developments concerning the paradox of information loss. Whilst
Hawking's calculations indicated that emitted radiation was thermal
\cite{Hawking:1974sw}, recent arguments have suggested that there are in fact
correlations between emitted quanta which can carry away information and therefore
preserve unitarity. The tunneling method of Parikh and Wilczek \cite{Parikh:1999mf}, 
in particular, has been an important milestone on this particular journey. There are
several puzzling issues which remain, even if the information paradox is resolved,
which pertain to the nature of the entropy on the horizon - and its corresponding
microstates. Such questions would hopefully be answered by a quantum theory of
gravity, however even within string theory such questions are often difficult
to pose unless we include supersymmetry. One particular class of stringy models that
may be exactly solvable are those of dilaton gravity in $1+1$ dimensions
in the type 0 string \cite{Gukov:2003yp, Berkovits:2001tg,Davis:2004xb}. 
Such a string only contains bosonic degrees of freedom due to the non-chiral
projection acting on the two-dimensional superstring. It has long been known that
charged (asymptotically flat) black hole solutions exist in this theory, and its
relative simplicity means that it is a good toy model for many of our current
conjectures regarding black hole information loss. Because the theory arises from a
truncation of the full superstring, and one can also use it to test matrix models
\cite{Olsson:2005en, Danielsson:2004xf}.

In this note we will be interested in the thermodynamic geometry of such a black hole
configuration, in particular we wish to understand whether a phase
transition should be expected at a critical value of the entropy as proposed in
\cite{Kim:2011fh}. We begin by briefly recapitulating the black hole solution before 
moving on to a discussion of information metrics and different bases in which they
are defined. 
\section{Black holes in type $0A$ string theory}
It has long been known that the low energy effective action for the type $0A$ theory,
with a single non-zero RR flux term and no tachyon, can be written as follows
\cite{Gukov:2003yp, Berkovits:2001tg,Davis:2004xb}
\begin{equation}
S = -\int d^2 \sigma \sqrt{-g}\left\lbrace e^{-2\Phi}\left(R + 4 (\nabla \Phi)^^2 +
4k^2 \right) + \Lambda \right\rbrace
\end{equation}
where $k^2= 2/ \alpha'$ is a cosmological constant term depending on the string
length, and $\Lambda=-Q^2/(2\pi \alpha')$ is a constant coming
from the gauge field via $F^{+}_{01}=-Q/(2\pi \alpha')$. Denoting the coordinates by
$\sigma = (t, \phi)$ we find there is a black hole solution
corresponding to the following;
\begin{equation}
ds^2 = - f(\phi) dt^2 + \frac{d\phi^2}{f(\phi)}, \hspace{0.5cm} \Phi = -k \phi
\end{equation}
where the function $f(\phi)$ has the following form
\begin{equation}
f(\phi) = 1 -\frac{1}{2k}(M-\Lambda \phi) e^{-2k\phi}
\end{equation}
with $M$ corresponding to the mass of the black hole. The event horizon exists as the
solution to
\begin{equation}\label{eq:master}
e^{-2 k \phi_H}(M-\Lambda \phi_H) = 2k
\end{equation}
and the temperature is given by the usual expression
\begin{equation}
T = \frac{f'(\phi)}{4\pi} \bigg|_{\phi_H} = \frac{1}{8k\pi}(\Lambda + 2k(M-\Lambda
\phi_H)) e^{-2 k \phi_H}.
\end{equation}
Using the WKB approximation, one can argue that Hawking radiation arises from quantum
tunneling of particles through the event horizon\footnote{We refer the reader to \cite{Kim:2011fh} for
further details.}. A simple calculation then yields the following result for the
temperature
\begin{equation}
T = \frac{k}{2\pi}\left(1+ \frac{\Lambda}{4 k^2}e^{-2k\phi_H} \right)
\end{equation}
and therefore one can use the standard thermodynamic relation $dM = T dS$ combined
with (\ref{eq:master}) to determine the entropy, which becomes
\begin{equation}
S = 4 \pi e^{2 k \phi_H}.
\end{equation}
Given the entropy of the system, one can then express the mass (in the fixed charge
ensemble) as a function of the entropy and the charge
\begin{equation}
M = \frac{k S}{2 \pi} - \frac{kQ^2}{8\pi} \ln \left( \frac{S}{4\pi}\right)
\end{equation}
allowing us to derive the following thermodynamic variables
\begin{eqnarray}\label{eq:thermo_variables}
T &=& \frac{\partial M}{\partial S} = \frac{k}{2\pi} \left(1-\frac{Q^2}{4S} \right),
\nonumber \\
\Phi &=& \frac{\partial M}{\partial Q} = - \frac{k Q}{4\pi}\ln \left(
\frac{S}{4\pi}\right), \nonumber \\
C &=& \frac{\partial M}{\partial T} = \frac{4S^2}{Q^2}-S, \nonumber \\
\chi &=& \frac{\partial Q}{\partial \Phi} = -\frac{4\pi}{\ln(S/4\pi)}.
\end{eqnarray}
The expression for the specific heat is interesting because it implies that (in
physically acceptable regions) the specific heat is \emph{positive} which is in stark
contrast to most black hole solutions. This suggests that the black hole is
thermodynamically stable, and suggests that the canonical  ensemble is valid. The
temperature of the solution is also bounded from above by $T_{max} = 1/ \sqrt{2 \pi^2
\alpha'}$ which corresponds to the Hagedorn temperature $T_H$. It is believed that
one can analytically continue to temperatures greater than $T_H$ using T-duality, in
which case one simply studies the black hole solution in type $0B$ at $T < T_H$
\cite{Olsson:2005en}. The expectation is that there will then be a phase transition
occurring at the Hagedorn temperature, which separates these two distinct phases. However
it is not clear whether one can study fluctuations through this point.


Note that the non-zero charge allows for the existence of an extremal black hole
solution, which can be found by demanding both $f(\phi_{ex}) = f'(\phi_{ex})=0$,
which has a mass 
\begin{equation}\label{eq:mex}
M_{ex} =  \frac{k Q^2}{8\pi}\left(1 - \ln \left( \frac{Q^2}{16 \pi}\right) \right).
\end{equation}
and zero-temperature (as expected for an extremal solution). Positivity of the
extremal mass therefore imposes an upper bound on the flux such
that $Q^2 \le 16 \pi e \approx 136.64$.
\subsection{The Phase Transition Puzzle}
In Davies paradigm~\cite{davies}, one can see that the specific heat (\ref{eq:thermo_variables}) does not diverge, 
therefore a phase transition is not expected. However if one makes a Legendre
transformation to work in the fixed potential ensemble, then the specific heat can be written in the
form
\begin{equation}
C_{\Phi} \sim \frac{S}{4\pi^2 \Phi^2} \ln \left( \frac{S}{4\pi}\right) \frac{4\pi^2
\Phi^2 - k^2 S \ln(S/4\pi)^2}{2+\ln(S/4\pi)}
\end{equation}
which diverges at $S = S_c  =4\pi e^{-2}$, therefore one may expect a phase
transition at this value of the entropy. Such phase transitions are commonly studied in the condensed matter community, but
less so in the black hole community. Since this
value is independent of any other variables, one may expect something unusual occur
at $S=S_c$ in the fixed charge ensemble but this is actually not the case.
All the thermodynamic variables are continuous through the curve $S=S_c$.
\begin{figure}
\centering
\includegraphics[scale=0.4]{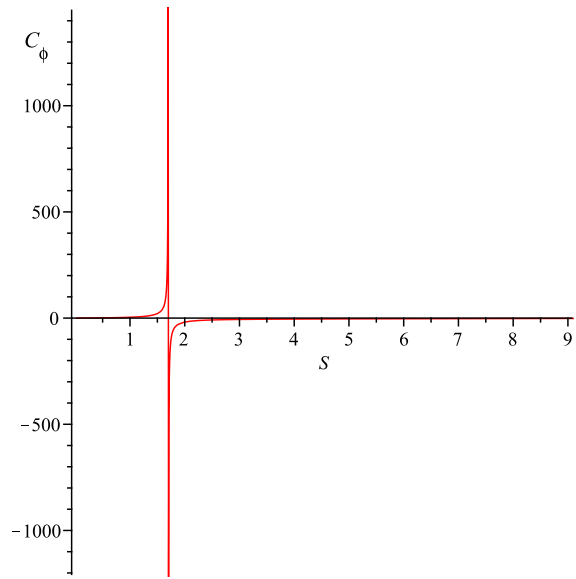}
\caption{Specific heat, $C_{\phi}$ , of the 0A black hole. The singular limit of the theory is the entropy cutoff, $S = S_c = 4\pi e^2$, which corresponds to the divergence of $C_{\phi}$. }
\end{figure}
One may also be concerned that we are not in Einstein frame, therefore the concept of
a gravitational phase transition may be ill defined.
However because our black hole solution is asymptotically flat, we can use the
results of \cite{Koga:1998un} to show that this is not the case.

In the following section we will study the black hole thermodynamics using information
geometry, which involves the construction of a metric
using the thermodynamic variables. One can then construct invariants from the metric,
which should be insensitive to the thermodynamic ensemble 
under consideration.
\section{Information Geometry}
Information geometry~\cite{IGBook} is the study of probability and information by way of
differential geometry. In this note we focus in particular on the Fisher type~\cite{ingemar} of
information geometry in that the Hessian matrix of the second derivatives of the
energy, or alternatively the entropy, can be regarded as a Riemannian metric on the
space of thermodynamical states. When energy is used as a potential this metric is
called the Weinhold metric~\cite{weinhold1}, when the entropy is used it is called
the Ruppeiner metric~\cite{ruppeiner1}.

Historically Ruppeiner proposed in 1979 a geometrical way of studying thermodynamics
of equilibrium systems. In his theory certain aspects of thermodynamics and
statistical mechanics of the system under consideration are encoded in a single
geometrical object, namely the metric describing its thermodynamic state space. The
distance on this
space is given by
\begin{equation}
ds^2_R = g^R_{ij} dX^i dX^j,
\end{equation}
where $g^R_{ij}$ is the so-called Ruppeiner metric defined as
\begin{equation}
g^R_{ij} = - \partial_i \partial_j S(X), \,\,\,  X = (U, N^a); \,\, a= 1,2, \ldots, n
\end{equation}
where $U$ is the system's internal energy (mass in our study), $S(X)$ is an 
entropy function of the thermodynamic system one wishes to consider and $N^a$ stand
for other extensive variables, or mechanically conserved charges of the system in 
our application. The minus sign in the definition is due to concavity of the entropy 
function. It was observed by Ruppeiner that in thermodynamic fluctuation theory the Riemannian 
curvature\footnote{Henceforth we refer to it as the Ruppeiner curvature.} of the Ruppeiner 
metric measures the complexity of the underlying statistical mechanical model, i.e. it is 
flat for the ideal gas whereas any curvature singularities are a signal of critical behavior. 
The Ruppeiner theory has been applied to numerous  systems and yielded significant results, 
for details see~\cite{ruppeiner2}. Ruppeiner originally developed his 
theory in the context of thermodynamic fluctuation theory, for systems in canonical ensembles.  
The Ruppeiner metric is conformally related to the Weinhold metric through
\begin{equation}
g^R_{ij} = \frac{1}{T} g^W_{ij}
\end{equation}
where $T$ is thermodynamic temperature of the system.  It is worth noting that some
flat Ruppeiner geometries can be accounted for by the flatness theorem~\cite{flatness}. 
More precisely, whenever the entropy function takes
the form
\begin{equation}
S = M^a f \left(\frac{M}{Q}\right),
\end{equation}
where $a \neq 1$ otherwise the Ruppeiner metric is degenerate. This theorem is, however, not applicable
to our study because of the form of the entropy.  The Ruppeiner method is not only usable for determining the underlying statistical 
mechanic model of the system, but also detecting its critical behavior. Incidentally it is consistent 
with the so-called {\it Poincar\'e}'s linear series method for analyzing stability in non-extensive systems. 
This method is simple owing to the fact that it utilizes only a few thermodynamic functions such as the fundamental 
relations in order to study/analyze (in)stabilities. This method can thus be applied to BHs although 
they are non-extensive systems. There have been some compelling results~\cite{poincare} but we will not 
dwell on this as it is outside the scope of our present work. 

Applications of this type of information geometry to black holes can be found in e.g.~\cite{stockholm1, 
withDaniel, Sarkar1, Myung1, ruppeinerBH1, Janke2010, Sahay2010, Gergely2011}. This method has also been 
applied to exotic systems such as the hot QCD model~\cite{hotQCD}. An alternate approach known as Geometrothermodynamics (GTD)
\cite{Quevedo:2007mj, Alvarez:2008wa} has been devoted to constructing Legendre invariant information metrics, which also appears to yield useful information about the black hole state space. This approach ensures that the form of the thermodynamic potential
does not dictate the phase space structure of the black hole. In this paper we will apply it to the study of the 0A black hole thermodynamics. 

Majority of black holes have negative specific heats (which are characteristics of self-gravitating systems) and are 
described microcanonically.  In spite of this we have found  that the Ruppeiner geometry of black holes is often surprisingly simple and elegant. Furthermore some  findings are physically suggestive in particular for ultra-spinning Myers-Perry black 
holes, in which the onset of ultra-spinning instability is detected by the singularity of the Ruppeiner
curvature~\cite{stockholm2}. 

\subsection{Weinhold analysis}
It is natural to work in the fixed charge ensemble, since this implies that the black
hole solution is in thermal \emph{and} electrical equilibrium with the surrounding heat bath.
\begin{equation}
dM = T dS + \Phi dQ,
\end{equation}
which ensures that we obtain the correct temperature for the black hole system. 
The Weinhold analysis can then be determined by identifying the mass of the black hole
with the internal energy. The Weinhold analysis then follows by determining the fluctuations of the mass
due to changes in the entropy-charge phase space, allowing one to construct a thermodynamic metric from
the corresponding Hessian. Note that when $S = 4\pi$, the term proportional to $Q^2$ in 
the definition of the mass will vanish (\ref{eq:thermo_variables}) and therefore 
our two-parameter system reduces to a one-parameter system. Physically one sees 
that $\Phi$ vanishes at $S = 4\pi$, therefore there is no source for the electric 
charge. For physical solutions we must then ensure $S \ne 4\pi$.

The line element for the Weinhold metric can be written in the form
\begin{equation}\label{eq:weinhold}
ds_w^2 = \frac{k}{8\pi}\left( \frac{Q^2}{S^2}dS^2 - 4 \frac{Q}{S} dS dQ - 2\ln\left(
\frac{S}{4\pi} \right)dQ^2 \right)
\end{equation}
which is conformally flat (as expected for a two-dimensional metric). 
To see this in detail first define the new variable $u=Q^2/S$ to eliminate the
cross-terms in the metric, allowing us to work in the
$u,Q$ basis. Then define a new variable
\begin{equation}
\label{eq:taudef}
\tau = \mp \sqrt{2} Q \sqrt{2+\ln \left( \frac{Q^2}{4 \pi u}\right)}
\end{equation}
which immediately implies
\begin{equation}\label{eq:weinhold_flat}
ds_w^2 = \frac{k}{8\pi} \frac{\tau^2}{Q^2} (d\tau^2 - dQ^2).
\end{equation}
Note that the entire line element vanishes when $\tau \to 0$ which corresponds to $S
= S_c = 4\pi e^{-2}$. This will give rise to a singularity in the
curvature invariant, however this does not imply a phase transition. This singular
point marks the break-down of the classical theory of thermodynamic
fluctuations, in much the same way as we expect the scale factor singularity in FRW
cosmology to correspond to a break-down of the classical
description of space-time. We propose that $S_c$ is actually the minimum (classical)
entropy one can associate to this black hole system,
and therefore physical entropy satisfies $S > S_c$. Furthermore note that the
Weinhold metric has Lorentzian signature with $SO(1,1)$ symmetry.
The metrics in Eq.~(\ref{eq:weinhold}, \ref{eq:weinhold_flat}) allow us to construct
the Weinhold curvature invariants
\begin{eqnarray}
R_w &=& -\frac{4\pi}{k Q^2} \left(2 + \ln \left( \frac{S}{4\pi}\right) \right)^{-2}
\nonumber \\
R^w_{ab}R_w^{ab} &=& \frac{8 \pi^2}{k^2 Q^4}\left(2 + \ln \left(
\frac{S}{4\pi}\right) \right)^{-4}
\end{eqnarray}
which are both singular at the solution $S_c$ indicating that this is indeed a
singular point of the thermodynamic metric.

\subsection*{Weinhold state space}
We can transform the metric in Eq.~(\ref{eq:weinhold_flat}) into a manifestly flat
form (in this case Minkowskian metric) in order
to study the state space of this BH and compare it to that of other BHs studied
elsewhere. 
We first write it in Rindler form ($ds^2 = -d\alpha^2 + \alpha^2 d\beta^2$) where we
obtain
\begin{equation}
\alpha = \sqrt{\frac{-k}{8\pi}}\frac{\tau^2}{2Q}, \hspace{0.5cm} \beta =
\frac{2Q}{\tau}.
\end{equation}
and then we transform this to Minkowski form using the hyperbolic transformations
\begin{equation}
t = \alpha \cosh \beta, \hspace*{.5cm} x = \alpha \sinh \beta.
\end{equation}
The opening of the wedge embedded in this Minkowskian null cone is given by 
\begin{equation}
\Delta = \frac{t}{x} = \coth \beta = \coth \left(\frac{2Q}{\tau}\right).
\end{equation}
With the definition of $\tau$ in Eq.~(\ref{eq:taudef}) we write this as a function of
the entropy
\begin{equation}
\Delta(S) = \coth \left(\frac{\sqrt{2}}{\sqrt{2 + \ln (S/4\pi)} }\right).
\end{equation}
Because the entropy has a cutoff limit at $S = S_c$ we see that $\beta$ diverges which implies that $\Delta = 1$.  
This becomes the singular limit for the 0A BH instead of $S=0$ for black holes we have studied previously. Indeed one can see
that the $\coth$ function is imaginary for $S < S_c$, and only becomes real (and smoothly increasing) for physical entropy
satisfying $S > S_c$.
\begin{figure}
\centering
\includegraphics[scale=1]{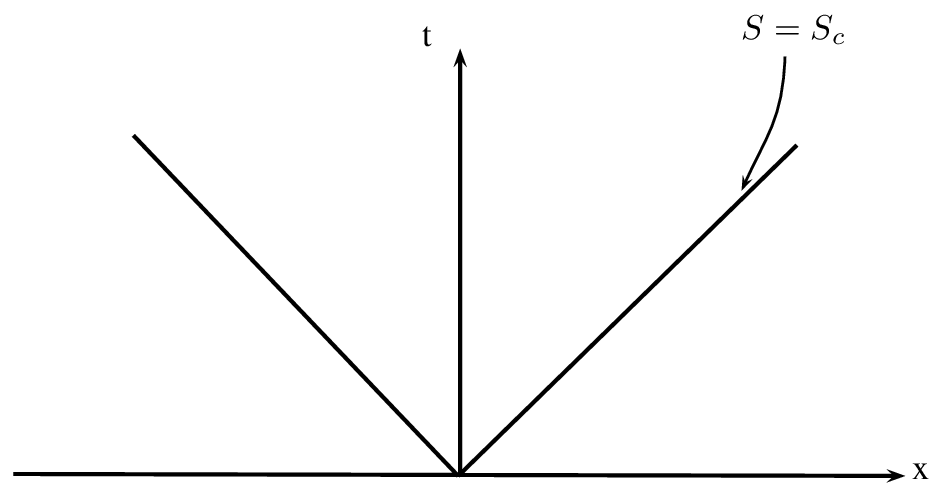}
\caption{A Weinhold state space of the 0A black hole. The null cone represents a 
singular limit which is the entropy cutoff in this case, $S = S_c = 4\pi e^2$. }
\end{figure}
 
To study the physical fluctuations, however, one must work in the Ruppeiner geometry
since the Ruppeiner scalar curvature is a measure of the correlation length of the system.
The metric can be calculated to give
\begin{equation}
ds_r^2 = \frac{\tau^2}{4 Q^2} \left(1 - \frac{Q^2}{4\pi} \exp
\left(\frac{4Q^2-\tau^2}{2Q^2}\right)\right)^{-1} (d\tau^2-dQ^2)
\end{equation}
which again has an apparent singularity as $\tau \to 0$ where the thermodynamic
description breaks down. 
The corresponding Ruppeiner curvature invariant becomes
\begin{equation}
R_r = \frac{3 Q^4+16 Q^2 S + 16S^2 + (Q^4+28 Q^2S) \ln(S/4\pi)+8Q^2S\ln(S/4\pi)^2}{2
Q^2(Q^2-4S)S (2+\ln(S/4\pi))^2}.
\end{equation}
which blows up at $S=S_c$ and also $S = Q^2/4$, which is the extremal limit of the
black hole.

There are many other approaches to information geometry which one could consider. Let
us therefore examine the GTD approach \cite{ Alvarez:2008wa} where
the metric is given by $g = M \partial_{ab}M dE^a dE^b$ where $E^a = {S,Q}$ which is
Legendre invariant.
The corresponding Weinhold invariant for this metric becomes
\begin{eqnarray}
R_w &=& -\left( \frac{64 \pi^2}{k^2 Q^2}\right) \frac{\cal{N}}{(4S-Q^2
\ln(S/4\pi))^3(2+\ln(S/4\pi))^2} \nonumber \\
\cal{N} &=& 2S[20S+Q^2(12+\ln(4\pi))]-2Q^2\ln(S)+8S^2\ln(S/4\pi) \nonumber \\
&+&Q^2(Q^2-10S)\ln(S/4\pi)^2-4Q^2S\ln(S/4\pi)^3
\end{eqnarray}
which appears to diverge along the curve spanned by $S = Q^2/4 \ln (S/4\pi)$, which
we denote by the solution $S_0$ 
\footnote{The solution is invertible, but its precise form is not relevant for our
purposes.}.
We argue that this is not a physical value for the entropy because $M(S_0) = 0$ and
there would be no black hole. However this is only an apparent singularity in the curvature invariant,  and one can
show that $R_w$ is in fact finite along this
curve. The only \emph{real} divergence of the above invariant is again at $S=S_c$.
We compute the Ruppeiner invariant in the usual manner
\begin{equation}
R_r = \frac{A(S,Q)}{B(S,Q)} \nonumber
\end{equation}
where we have defined
\begin{eqnarray}
\cal{A} &=& -(4 \pi (4 S (128 Q^2 S^2 + 320 S^3 + Q^6 (12 + 5 \ln(4 \pi)) + 4 Q^4 S
(-17 + 8 \ln(4 \pi))) \nonumber \\
&-& 4 Q^4 S (5 Q^2 + 32 S) \ln(S) + \ln(S/4 \pi) (256 S^3 (Q^2 + S) + Q^2 \ln(S/4
\pi) (4 (Q^6 - 5 Q^4 S - 8 Q^2 S^2 - 48 S^3) \nonumber \\
&+& \ln(S/4 \pi) (Q^6 + 20 Q^4 S - 128 S^3 + 8 Q^4 S \ln(S/4 \pi)))))) \nonumber \\
\cal{B} &=& (k Q^2 (Q^2 - 4 S) S (2 + \ln(S/4\pi))^2 (-4 S + Q^2 \ln(S/4 \pi))^3).
\end{eqnarray}
which, like before, has divergences only at $S=S_{ex}, S=S_c$.

There are, however, a large class of Legendre invariant metrics available (in this
basis) and we should consider that they may lead to different results. Two common metrics
can be  parametrized as follows
\begin{eqnarray}
g^w &=& (S \partial_S M + \epsilon Q \partial_Q M) \rm{diag}(-M_{SS}, M_{QQ}) 
\end{eqnarray}
where the first metric corresponds to $\epsilon =1$ and the second one to
$\epsilon=0$. 
The Ruppeiner curvature invariant can be computed in both cases with the result
\begin{eqnarray}
R^{\epsilon=1}_r &=& \frac{32\pi^2}{k^2 Q^2 \ln(S/4\pi)^2[Q^2-4S+2Q^2 \ln(S/4\pi)]^3}
\left((Q^2-4S)^2 -16 S^2 \ln(S/4\pi) \right) \\
&+& \frac{64\pi^2}{k^2\ln(S/4\pi)[Q^2-4S+2Q^2 \ln(S/4\pi)]^3}
\left(Q^2+6S+4\ln(S/4\pi)(Q^2+2S+2S\ln(S/4\pi)) \right) \nonumber \\
R^{\epsilon=0}_r &=& \frac{32 \pi^2}{k^2 Q^2} \frac{(Q^2-4S)^2+16S^2
\ln(4\pi)+4S(-4S\ln(S)+Q^2\ln(S/4\pi)[S+2\ln(S/4\pi)])}{(Q^2-4S)^3 \ln(S/4\pi)^2}
\end{eqnarray}
which only has singularities at $S=S_{ex}$. This is interesting because the
singularity at $S_c$ has been removed in this picture
\subsection{The Ruppeiner analysis}
The Weinhold method is useful because the Hessian is easily calculated, but does it
contain all the information about the theory?
Let us consider this question by working directly with the Ruppeiner analysis, where the
information metric can be considered as the pull-back
of the entropy function to a manifold spanned by \emph{physical} observables
$(M,Q)$. Indeed this is the form of the metric as first proposed by Ruppeiner from
thermodynamic fluctuation theory.
The entropy clearly takes the form
\begin{equation}
S(M,Q) = -\frac{Q^2}{4} W \left \lbrack -\frac{16 \pi}{Q^2} e^{-8\pi M/kQ^2}
\right\rbrack
\end{equation}
where $W[z]$ is the Lambert-W function, which we will often employ as short-hand
notation in what follows. It is straightforward to show that the extremal solution 
occurs when $S=Q^2/4$, corresponding to the condition
\begin{equation}
M_{ex} = -\frac{k Q^2}{8 \pi} \ln \left( \frac{Q^2}{16 \pi e} \right)
\end{equation}
which is identical to $(\ref{eq:mex})$. The Ruppeiner metric in this case is determined via the
Hessian 
$$g_{ij}= -\partial_i \partial_j S(M,Q),$$ 
and the scalar curvatures are shown to be of the form
\begin{eqnarray}
R_r &=& \frac{2 W[z]{(16\pi M-5kQ^2)(8\pi M -kQ^2)+2kQ^2W[z](16\pi
M-4kQ^2+kQ^2W[z])}}{W[z](1+W[z])(8\pi MQ-2kQ^3+kQ^3W[z])^2} \nonumber \\
&+& \frac{2kQ^2[8\pi M -3kQ^2]}{W[z](1+W[z])(8\pi MQ-2kQ^3+kQ^3W[z])^2} \nonumber \\
R_w &=& -\frac{4k \pi Q^2}{(8\pi M-2k Q^2 + kQ^2 W[z])^2}
\end{eqnarray}
The Ruppeiner invariant has a single divergence at $M_{ex}$ as expected. The Weinhold
invariant appears to diverge at
\begin{equation}
M_c = \frac{8 k\pi + e^2 kQ^2}{4 \pi e^2}
\end{equation}
but the Ruppeiner invariant remains finite at this value. One can actually identify
$M_c$ with $S_c$ in the Weinhold picture, which we 
believe to be a cut-off for the entropy, therefore the Ruppeiner invariant is
insensitive to this value. This suggests that the 
black hole must have mass greater than $M_C$ for the classical theory to be valid.

Let us now consider what happens to the GTD theory in the Ruppeiner analysis. In this
case the metric takes the form
\begin{equation}
g = -M \left( \frac{\partial S}{\partial M}\right)^{-1} \frac{\partial^2 S}{\partial
X^a \partial X^b} dX^a dX^b
\end{equation}
which results in the Ruppeiner curvature invariant
\begin{eqnarray}
R_r &=& \frac{1}{8M^3}\left(-4M+\frac{kQ^2}{\pi} + \frac{2}{\pi}(4M\pi + kQ^2)W[z] +
\frac{kQ^2}{\pi}W[z]^2 + \frac{8kM Q^2(kQ^2-12 M \pi)}{(8M\pi -2kQ^2+kQ^2 W[z])^2}
\right) \nonumber \\
&+&\frac{1}{4 M^3 \pi} \frac{(16 M^2 \pi^2-6Mk\pi Q^2+k^2Q^4)}{(8\pi M -2kQ^2+kQ^2
W[z])}
\end{eqnarray}
where we again employ the short handed notation for the Lambert function. This
invariant does not have a non-trivial divergence 
\footnote{Trivial in this sense refers to setting one of the parameters to zero.},
which is somewhat troubling because we expect the solution
to diverge at extremality.

Let us summarize the results in the following table, where we list all the
divergences of the Ruppeiner invariant.
\begin{table}[htp]
\begin{center}
\begin{tabular}{|c|c|c|}
\hline
Formalism & Weinhold & Ruppeiner \\ \hline
Standard & $S_c$,  $S_{ex}$ & $M_{ex}$ \\ \hline
GTD & $S_c$, $S_{ex}$ & none \\ \hline
$\epsilon$ GTD & $S_{ex}$ & - \\
\hline
\end{tabular}
\end{center}
\caption{Divergences of Ruppeiner invariant}
\end{table}
Without a detailed analysis one would simply conclude that in the Weinhold analysis
there was a possible phase transition at $S=S_c$, however we 
believe that the classical thermodynamic picture breaks down at this value, rather
than suggesting the appearance of a phase transition.
The underlying physics behind such a cut-off is not apparently clear because the
thermodynamic observables
are smoothly varying functions. However the fact that there is no additional
dependence on the charge suggests that $S_c$ is a fundamental
scale of the theory, and the notion of equilibrium fluctuations breaks down at this
point.

In the Ruppeiner analysis only the physical divergences seem to appear ie there are no
signs of divergent behaviour at $S_c$. However the GTD approach does not include the
extremal case, suggesting that only the Ruppeiner invariant
in the Ruppeiner analysis is capable of computing all the relevant physical information.
The $\epsilon$ GTD description (in the Weinhold analysis) also
only selects this singular point.

\section{Discussion}
In this note we have considered the case of possible phase transitions in the $OA$
black hole solution using information geometry. 
The potential phase transition in the fixed potential ensemble turns out to be
associated with a singularity in the description of
the thermodynamic geometry, where the line element vanishes. We interpret this as a
(hard) cut-off for the entropy of the black hole, which
must satisfy $S > S_c$ in the physical domain - although we do not have a compelling
thermodynamic explanation for its existence. We believe that this singularity represents the 
break-down of equilibrium thermodynamics, rather than indicating a phase transition. The distance 
between equilibrium fluctuations is squeezed as we approach the $S_c$ curve, until it becomes ill defined
precisely at $S=S_c$ indicating that all fluctuations exist at the same singular point.

Moreover we compared and contrasted several approaches to the problem in an attempt
to understand which curvature invariants contain the 
physical information about the thermodynamic ensemble. Our results suggest that one
should work predominantly with the Ruppeiner analysis, since 
the invariants in the Weinhold analysis tend to pick out all divergences - which may
not all correspond to phase transitions. Indeed this is natural from the perspective of
equilibrium thermodynamic fluctuation theory.

Furthermore the Ruppeiner invariant, in the Ruppeiner analysis, captures the physically 
relevant divergences (including the extremal point). 
Our results are preliminary and more detailed comparison of the two approaches is
likely to yield better understanding of the power
of contact geometry. It would also be useful to understand the physical relevance of the sign of the 
curvature invariant, since this should also tell us something useful about the microstates of the
black hole theory. 

\subsection*{Acknowledgements}

Narit Pidokrajt is grateful for the hospitality of the KoF group, Stockholm University where most of this research work is conducted.  We would like to thank Ingemar Bengtsson, Hans Hansson and George Ruppeiner for informative discussions. NP is 
financially supported in part by grant number FOA10V-116 from the Royal Swedish Academy of Sciences, and from the Helge Ax:son Johnson Stiftelse of Sweden.  JW is supported in part by NSERC of Canada.

\end{document}